\RequirePackage{fix-cm}
\documentclass[twocolumn]{svjour3}          
\smartqed  
\usepackage{graphicx}
\usepackage{amsmath,amssymb}
\usepackage{mathptmx}      
\usepackage{hyperref} 
\newcommand\nn{\nonumber\\}
\newcommand\beq{\begin{equation}}
\newcommand\eeq{\end{equation}}
\newcommand\bea{\begin{eqnarray}}
\newcommand\eea{\end{eqnarray}}

\newcommand\mpc{{\rm Mpc^{-1}}}
\newcommand\kms{{\rm km s^{-1}}}

\newcommand\ie{{\rm i.e.}}
\newcommand\eg{{\rm e.g.}}
\newcommand\fig{{\rm Fig.}}
\newcommand\bwt{\begin{widetext}}
\newcommand\ewt{\end{widetext}}

\newcommand\LCDM{$\Lambda${\rm CDM}}
 
\newcommand\dsum{\displaystyle\sum}
\begin{document}
\title{Dark energy homogeneity in general relativity: Are we applying it correctly?}
\author{Didam G.A. Duniya}
\institute{	Didam G.A. Duniya \at Physics Department\\ University of the Western Cape\\ Cape Town 7535, South Africa\\ \email{adamsgwazah@gmail.com} }

\date{Received: date / Accepted: date}

\maketitle
\begin{abstract}
Thus far, there does not appear to be an agreed (or adequate) definition of homogeneous dark energy (DE). This paper seeks to define a valid, adequate homogeneity condition for DE. Firstly, it is shown that as long as $w_x \neq -1$, DE must have perturbations. It is then argued, independent of $w_x$, that a correct definition of homogeneous DE is one whose density perturbation vanishes in {\em comoving} gauge: and hence, in the DE rest frame. Using phenomenological DE, the consequence of this approach is then investigated in the observed galaxy power spectrum -- with the power spectrum being normalized on small scales, at the present epoch $z=0$. It is found that for high magnification bias, relativistic corrections in the galaxy power spectrum are able to distinguish the concordance model from both a homogeneous DE and a clustering DE -- on super-horizon scales. 
\end{abstract}

\section{Introduction}\label{sec:intro}%
Dark energy (DE) is {\em dark}, but the underlying physics of DE is even {\em darker}. Understanding the nature of DE remains a puzzle in general relativity. A long standing question is that: is DE actually static vacuum energy $\Lambda$, \ie~like in the standard concordance model (\LCDM); or a dynamic field, \eg~like in the quintessence \cite{Duniya:2013eta}--\cite{Brax:2000yb} models (QCDM)? If DE is described by $\Lambda$ then it can not have perturbations (or evolve) at all. Although \LCDM\ is the best-fit model to the current data, other alternatives have been considered in the literature, \eg~a homogeneous dynamical DE. However, if DE is dynamical then it can have perturbations. How then do we define a valid homogeneous dynamical DE? 

There does not appear to be an agreed or adequate definition of homogeneous (dynamical) DE. For example, if the DE physical sound speed is $c_{sx} = 1$, then DE cannot cluster on sub-Hubble scales. Thus it is sometimes concluded that DE is approximately homogeneous (see \eg~\cite{Takada:2006xs}--\cite{Piattella:2014lba}). The caveat to this assumption is that it only ensures a `scale-dependent' homogeneity, in the sense that it makes DE homogeneous only on sub-Hubble scales, but on super-Hubble scales DE becomes significantly inhomogeneous. This is because, $c_{sx}=1$ implies that the DE density perturbations propagate with the speed of light; hence DE fails to cluster, and is perturbatively insignificant. However, on (Hubble) horizon scales the perturbation behaviour is different and the homogeneity assumption breaks down, \ie~the implicit assumption of no clustering in DE on super-horizon scales, given that $c_{sx}=1$, is inconsistent.

Moreover, an assumption often used for DE homogeneity is the requirement that all its perturbations vanish~\cite{Duniya:2013eta}--\cite{Devi:2014rwa}, \ie~by setting the DE density perturbation and velocity potential to (absolute) zero~\cite{Duniya:2013eta}:
\beq\label{falseHomDE}
\delta_x \;=\; 0 \;=\; V_x,
\eeq
where the associated evolution equations are therefore discarded. However, it has been pointed out that a fluctuating, inhomogeneous component is the only valid way of introducing an additional energy component (\ie~DE): a smooth (non-fluctuating), time-varying component is unphysical -- it violates the equivalence principle \cite{Caldwell:1997ii}. Moreover, it is known that Eq. \eqref{falseHomDE} leads to a violation of the self-consistency of the equations of general relativity, by causing a contradiction in the equations. For example, it has been shown that Eq. \eqref{falseHomDE} leads to a false boost in the matter power spectrum~\cite{Duniya:2013eta,Hwang:2009zj}, and in the integrated Sachs-Wolfe effect~\cite{Jassal:2012pd,Caldwell:1997ii} (in the cosmic microwave background) -- on horizon scales. (Also, recently it has been shown in \cite{Nesseris:2015fqa} that neglecting the DE perturbations can lead to misleading estimation of the matter growth index, giving up to ${\sim}\,3\%$ deviation: which is a significant amount, as we enter an era of precision cosmology.) Nevertheless, none of the previous works has shown explicitly what the inconsistency resulting from Eq. \eqref{falseHomDE} is, nor has given any suggestions on how to solve or circumvent this problem.

In this work we show analytically the inconsistency resulting from Eq. \eqref{falseHomDE}. We propose an alternative way, via the intrinsic entropy perturbation, to define a suitable condition for DE homogeneity -- which corrects Eq. \eqref{falseHomDE} to avoid the violation of the consistency of the equations of general relativity, and also eliminates $c_{sx}$ from the equations (like in \LCDM). It should be noted that the aim of the paper is not to fit the given homogeneous DE to the data (which might have been one other avenue to show the observational significance of the homogeneous DE), but to provide a suitable way to define a valid, adequate DE homogeneity condition -- which currently seems to be non-existent. Furthermore, for illustration purpose, the effects of general relativistic corrections (and magnification bias) in the galaxy power spectrum is demonstrated.

We begin by outlining the basic background equations in section~\ref{sec:bkgnd}; we give the perturbations equations in section~\ref{sec:Perts}. In section~\ref{sec:Entropy} we discuss the intrinsic entropy perturbation; and in section~\ref{sec:GRhomDE} we look at DE homogeneity in general relativity: discussing the `unphysical' smooth DE and a `true' homogeneous DE -- illustrating their effects in the galaxy and matter power spectra. We conclude in section~\ref{sec:Concln}.

\section{The Background Equations}\label{sec:bkgnd}
The standard acceleration equation (see e.g. \cite{Kurki-Suonio:2015cpt}) is:
\beq\label{HEvoln}
{\cal H}' = -\dfrac{1}{2} {\cal H}^2 (1+3w),\quad w \;\equiv\; \dsum_A{\Omega_A{w}_A},
\eeq
where ${\cal H}=a'/a$ is the comoving Hubble parameter, prime denotes derivative with respect to conformal time, with $a$ being the scale factor; $\Omega_A = \bar{\rho}_A / \bar{\rho}$ is the energy density parameter (with over bars denoting background) of species $A$, which evolves according to the equation
\beq\label{OmegaEvoln}
\Omega_A' = -3{\cal H}(w_A - w)\Omega_A,
\eeq
with $\bar{\rho}_A$ being the background energy density of species $A$, and $\bar{\rho}$ is the total energy density of all the species. Similarly, $w_A = \bar{p}_A / \bar{\rho}_A$ is the equation of state parameter of species $A$, which evolves by 
\beq\label{EoSEvoln}
w_A' = -3{\cal H}(1+w_A)(c^2_{aA} - w_A), 
\eeq
where $c^2_{aA} =\bar{p}_A' / \bar{\rho}_A'$ is the adiabatic sound speed associated with species $A$, and $\bar{p}_A$ is the background pressure.

\section{The General Perturbations Equations}\label{sec:Perts}%
Here we adopt the Newtonian metric, given by
\beq\label{metric}
ds^2 = a^2\left[-\left(1+2\Phi\right) d\eta^2 + \left(1-2\Phi\right)d\vec{x}^2\right],
\eeq
where $\eta$ is the conformal time, and $\Phi$ is the (Newtonian) gravitational potential. Note that by the choice of the metric~\eqref{metric} we assume zero (or negligible) anisotropic stress: this assumption is crucial for the subsequent derivations. The relativistic Poisson equation is given by
\beq\label{PoissonEq}
\nabla^2\Phi = \dfrac{3}{2} {\cal H}^2 \dsum_A{\Omega_A\Delta_A},
\eeq
where the comoving density perturbation $\Delta_A$ is given by
\beq\label{DeltaDefn}
\Delta_A \;\equiv\; \delta_A + \dfrac{\bar{\rho}'_A}{\bar{\rho}_A} V_A \;=\; \delta_A - 3{\cal H}(1+w_A)V_A,
\eeq
where $\delta_A = \delta{\rho}_A /\bar{\rho}_A$ and $\delta{\rho}_A$ is the energy density perturbation. The gravitational potential is driven by the total momentum density, given by
\beq\label{PhiEvoln}
\Phi' +{\cal H}\Phi = -\dfrac{3}{2} {\cal H}^2(1+w)V,
\eeq
where the 4-velocities are given by \cite{Duniya:2015nva}
\bea\label{overDens:Vels}
u^\mu_A = a^{-1}\left(1 -\Phi,\, \partial^i V_A\right), \quad u^\mu = a^{-1}\left(1 -\Phi,\, \partial^i V\right),
\eea
with $u^\mu$ being the total $4$-velocity and  $V$ is the total velocity potential, given by 
\beq\label{totalV}
V = \dfrac{1}{1+w}\dsum_A{\Omega_A\left(1+w_A\right)V_A}.
\eeq
  
We consider all species as perfect fluids. Thus for the species $A$, the perturbed energy-momentum tensor is 
\bea\label{pertEMT}
\delta{T}^{\mu\nu}_A &=& \left(\delta\rho_A +\delta{p}_A\right)\bar{u}^\mu_A\bar{u}^\nu_A +\delta{p}_A\bar{g}^{\mu\nu} + \bar{p}_A\delta{g}^{\mu\nu}\nn
&& + \left(\bar{\rho}_A + \bar{p}_A\right)\left[\delta{u}^\mu_A\bar{u}^\nu_A + \bar{u}^\mu_A\delta{u}^\nu_A\right],
\eea
where $\delta{p}_A$, $\delta{u}^\mu_A$ and $\delta{g}^{\mu\nu}$ are the perturbations in the pressure, $4$-velocity and the metric tensor, respectively. The conservation of energy and momentum implies that
\beq\label{pertConsvn}
\nabla_{\mu}\dsum_A\delta{T}^{\mu\nu}_A \;=\; 0 \;=\; \nabla_{\mu}\delta{T}^{\mu\nu}_A ,
\eeq
where the second equality follows from the assumption that the individual fluid species do not interact directly with one another: they only interact (indirectly) gravitationally via the Poisson equation~\eqref{PoissonEq}.

Thus given Eq. \eqref{pertConsvn} the velocity potential $V_A$ and the comoving overdensity $\Delta_A$ \eqref{DeltaDefn} evolve by
\bea\label{VelEvoln}
V_A' + {\cal H}V_A &=& -\Phi - \dfrac{c^2_{sA}}{1+w_A}\Delta_A, \\ \label{DeltaEvoln}
\Delta_A' - 3{\cal H}w_A\Delta_A &=& {\cal H}\hat{\Delta}_A -(1+w_A)\nabla^2{V}_A,
\eea
where we have defined the parameter $\hat{\Delta}_A$ by
\bea\label{DeltaHat}
{\cal H}\hat{\Delta}_A \;\equiv\; \dfrac{9}{2} {\cal H}^2(1+w_A)\dsum_{B\neq A}{\Omega_B(1+w_B)(V_A - V_B)}.
\eea
The index $B$ runs through the entire species, for a given (fixed) value of $A$.

\section{The Intrinsic Entropy Perturbation}\label{sec:Entropy}
The entropy of a given (thermodynamic) system or fluid, measures the degree of `disorderliness' of that fluid; hence is a perturbed quantity. The intrinsic (or inherent) entropy perturbation $\delta{s}_A$ of $A$, may be given by~\cite{Kodama:1985bj}--\cite{Bean:2003fb}
\beq\label{EntroPert} 
\bar{p}_A\,\delta{s}_A \;\equiv\; \bar{p}'_A\left(\dfrac{\delta{p}_A}{\bar{p}'_A} - \dfrac{\delta{\rho}_A}{\bar{\rho}'_A}\right),
\eeq
i.e. the entropy perturbation quantifies the part of the (effective) pressure perturbation that is not simply related to the (effective) energy density perturbation. Then the physical sound speed $c^2_{sA}$ of species $A$, is defined in the rest frame (``rf") of $A$ -- given by
\beq\label{cs2}
c^2_{sA} \;\equiv\; \left. \dfrac{\delta{p}_A}{\delta{\rho}_A} \right|_{\rm rf} \; \geq\; 0,
\eeq
where we note that this is essentially the speed of propagation of the pressure perturbation $\delta{p}_A$ relative to the density perturbation $\delta{\rho}_A$ -- when $A$ is at rest. Then by changing from some arbitrary frame $x^\mu$ into the rest frame $x^\mu|_{\rm rf}$, given by $x^\mu \rightarrow x^\mu|_{\rm rf} = x^\mu + \xi^\mu$, this leads to a gauge transformation of the energy-momentum tensor:
\beq\label{EMTens}
T^\mu_A\/_\nu \;\rightarrow\; \left. T^\mu_A\/_\nu \right|_{\rm rf} \;=\; T^\mu_A\/_\nu - {\cal L}_\xi \bar{T}^\mu_A\/_\nu,
\eeq
where the Lie derivative ${\cal L}_\xi$, with respect to the transformation $4$-vector $\xi^\mu$, is given by
\beq\label{LieDeriv}
{\cal L}_\xi \bar{T}^\mu_A\/_\nu \;=\; \xi^\alpha \partial_\alpha \bar{T}^\mu_A\/_\nu + \bar{T}^\mu_A\/_\alpha \partial_\nu {\xi}^\alpha - \bar{T}^\alpha_A\/_\nu \partial_\alpha {\xi}^\mu ,
\eeq
with $|\xi^\mu| \ll 1$. Thus the ($0$-$0$)th and the ($i$-$j$)th components of the transformation~\eqref{EMTens} yield, respectively 
\beq\label{rfRhoP}
\delta{\rho}_A|_{\rm rf} = \delta{\rho}_A -\xi^{0}\bar{\rho}_A',\quad \delta{p}_A|_{\rm rf} = \delta{p}_A -\xi^{0}\bar{p}_A',
\eeq
and the ($i$-$0$)th or ($0$-$j$)th component yields
\beq\label{rfVelB}
V_A|_{\rm rf} = V_A+\xi^0,
\eeq
with the velocity potential $V_A$ being given in the Newtonian gauge -- where it is automatically gauge-invariant. (One advantage of using the conformal Newtonian metric is that the resulting gravitational perturbations are automatically gauge-invariant.) However, when $A$ is at rest, we have
\beq\label{delTi0}
\left. T^0_A\/_j \right|_{\rm rf} = 0 = \left. T^i_A\/_0 \right|_{\rm rf},
\eeq
where it follows that $V_A|_{\rm rf} =0$, and thus $\xi^0 = -V_A$. Putting this in Eq.~\eqref{rfRhoP}, we then obtain by using Eq.~\eqref{cs2} that
\beq\label{deltaP}
\delta{p}_A = c^2_{aA}\delta{\rho}_A + (c^2_{sA} - c^2_{aA})\bar{\rho}_A \Delta_A,
\eeq
where $\Delta_A$ is given by Eq.~\eqref{DeltaDefn}. Whence we obtain the intrinsic entropy perturbation~\eqref{EntroPert}, given by
\beq\label{EntroPert2}
\bar{p}_A\,\delta{s}_A = (c^2_{sA} - c^2_{aA})\bar{\rho}_A \Delta_A,
\eeq
where $\Delta_A$ is gauge-invariant, and consequently so is $\delta{s}_A$. 
 
Moreover, given Eqs. \eqref{DeltaDefn} and \eqref{rfRhoP}--\eqref{delTi0}, we get
\beq\label{comvD}
\delta{\rho}_A|_{\rm rf} \;=\; \bar{\rho}_A \Delta_A .
\eeq
This implies that the comoving density perturbation of any species corresponds to the density perturbation of that species in its rest frame. Moreover, given Eq. \eqref{comvD}, we have Eq. \eqref{EntroPert2}: $\bar{p}_A\delta{s}_A = (c^2_{sA} - c^2_{aA})\, \delta{\rho}_A|_{\rm rf}$ -- which therefore implies that the `intrinsic entropy' of any species corresponds to the `entropy perturbation in the rest frame' of the species.

\section{General Relativity and Dark Energy Homogeneity}\label{sec:GRhomDE} %
In this section, we analytically discuss the inconsistency in the equations of general relativity -- resulting from Eq.~\eqref{falseHomDE}, we illuminate what the inconsistency really is: (analytically) describing its source/origin. We then propose a suitable way to define a valid, adequate condition for DE homogeneity in general relativity.

It is known that the equations of general relativity form a complete and consistent system. Thus an implication of this is that the gravitational potential evolution equation~\eqref{PhiEvoln} should always reduce to the Poisson equation~\eqref{PoissonEq}. To confirm this, it is only sufficient to show that the Poisson equation at any time solves the associated gravitational potential evolution equation. 

Hence by taking the time derivative of Eq.~\eqref{PoissonEq}, and using Eqs.~\eqref{HEvoln},~\eqref{OmegaEvoln},~\eqref{DeltaDefn}--\eqref{DeltaHat}, we get
\beq\label{nabla2PhiEvoln}
\nabla^2\Phi' = -{\cal H}\nabla^2{\Phi} - \dfrac{3}{2} {\cal H}^2 \dsum_A{\Omega_A(1+w_A)\nabla^2{V}_A},
\eeq
where by applying the inverse Laplacian to both sides, we get Eq.~\eqref{PhiEvoln} -- as required. This way, the system of equations remains complete and consistent. Nevertheless, note that Eq.~\eqref{nabla2PhiEvoln} is obtained mainly as a result of the fact that
\beq\label{Consistency}
\dsum_A{\Omega_A\hat{\Delta}_A} = 0,
\eeq
where given Eq.~\eqref{DeltaHat}, it is easy to establish Eq.~\eqref{Consistency}. 

Equation~\eqref{Consistency} is essentially the statement of a `consistency condition' for the system of equations of general relativity. This condition should always hold given any correct set up -- within general relativity. (Note that Eq.~\eqref{Consistency} immediately holds true for \LCDM. However, as will be shown subsequently, there are situations for dynamical DE where this condition would (i) hold true, and (ii) not hold true.)

However, if  in any adopted framework we have that this condition does not hold, \ie~$\sum_A{\Omega_A\hat{\Delta}_A} \neq 0$, then this will result in a contradiction: where we are unable to recover the standard gravitational potential evolution equation~\eqref{PhiEvoln}; effectively we will rather have a transformation, given by
\beq\label{PhiEvolnTrans}
\Phi' \;\rightarrow\; \Phi' + {\cal H}\hat{\Phi},
\eeq
where $\Phi'$ is given by Eq.~\eqref{PhiEvoln}, and 
\beq\label{PhiHat}
{\cal H}\hat{\Phi} \;\equiv\; \dfrac{3}{2} {\cal H}^3 \nabla^{-2} \dsum_A{\Omega_A\hat{\Delta}_A}.
\eeq
Thus Eqs.~\eqref{PhiEvolnTrans} and \eqref{PhiHat} analytically express the `unwanted' inconsistency (or contradiction) that will arise in the physical equations of general relativity, when Eq.~\eqref{Consistency} fails to hold. It should be noted that the parameter ${\cal H}\hat{\Delta}_A$ is physical, and will contribute to the comoving density perturbation via Eq.~\eqref{DeltaEvoln}. 

In the following subsections, we give the particular evolution equations and present a new definition for a `true' homogeneous DE.

\subsection{\bf The particular perturbations equations}\label{subsec:partlr}%
We assume (henceforth) the late Universe -- dominated by DE and matter ($``m"$), \ie~baryons and cold dark matter. Thus the relativistic Poisson equation is
\beq\label{Poisson}
\nabla^2\Phi = \dfrac{3}{2} {\cal H}^2 \left[\Omega_m\Delta_m + \Omega_x\Delta_x\right],
\eeq
where we take care to use the correct overdensities $\Delta_{m,x}$, in order to avoid `unphysical artefacts' (see \eg~\cite{Dent:2008ia}) in the results. 

The gravitational potential evolves by
\beq\label{Phiprime}
\Phi' +{\cal H}\Phi = -\dfrac{3}{2} {\cal H}^2\left[\Omega_m V_m + \Omega_x(1+w_x)V_x\right],
\eeq
and the matter perturbations evolve according to
\bea\label{VmEvoln}
V_m' + {\cal H}V_m &=& -\Phi, \\ \label{DmEvoln}
\Delta_m' -\dfrac{9}{2} {\cal H}^2\Omega_x(1+w_x)(V_m - V_x) &=& -\nabla^2{V}_m .
\eea
Similarly, the DE perturbations evolve by
\bea\label{VxEvoln}
V_x' + {\cal H}V_x &=& -\Phi - \dfrac{c^2_{sx}}{1+w_x}\Delta_x, \\ \label{DxEvoln}
\Delta_x' - 3{\cal H}w_x\Delta_x &=& \dfrac{9}{2} {\cal H}^2\Omega_m(1+w_x)(V_x - V_m)\nn
&& -(1+w_x)\nabla^2{V}_x,
\eea
where Eqs.~\eqref{Poisson}--\eqref{DxEvoln} follow from Eqs.~\eqref{PoissonEq}--\eqref{DeltaHat}. 

Thus Eq.~\eqref{falseHomDE} implies that $\Delta_x=0=\Delta'_x$, and Eq.~\eqref{DxEvoln} yields: $-(9/2) {\cal H}^2 \Omega_m(1+w_x)V_m = 0$, which implies that either (i) $w_x=-1$, or (ii) $V_m = 0$. But it is already taken that $w_x \neq -1$, and $V_m$ cannot be zero (in the given gauge). Hence this leads to a contradiction. Usually in the literature, the $\Delta'_x$ equation is merely disregarded (while keeping $w_x \neq -1$ and $V_m \neq 0$) -- this is the source of inconsistency in the general relativistic equations, which has rightly been reported in the literature (see \eg~\cite{Duniya:2013eta,Hwang:2009zj}). 

Hence given that $w_x \neq -1$, and since by Eq.~\eqref{falseHomDE} the $\Delta'_x$ equation (and hence, $\hat{\Delta}_x$) is discarded, we have
\bea\label{DmHat}
\dsum_{A=m,x}{\Omega_A\hat{\Delta}_A} = \dfrac{9}{2} {\cal H}^2\Omega_m\Omega_x(1+w_x)V_m = {\cal H}\Omega_m\hat{\Delta}_m,
\eea 
which eventually leads to Eq.~\eqref{PhiEvolnTrans}; thereby defying Eq.~\eqref{Consistency}. Obviously by Eq.~\eqref{DmHat} (and preceding explanations), unless $w_x = -1$, Eq.~\eqref{falseHomDE} leads to a contradiction -- and hence a violation of the self-consistency of the equations of general relativity. Thus Eq.~\eqref{falseHomDE} is wrong, and the resulting `smooth' DE is `unphysical'. However, Eq.~\eqref{DmHat} reveals why the \LCDM\ satisfies general relativity despite the fact that all the DE perturbations therein become zero: $w_x = -1$ in \LCDM. Basically, provided $w_x \neq -1$, DE must cluster.

\begin{figure}\centering%
\includegraphics[scale=0.4]{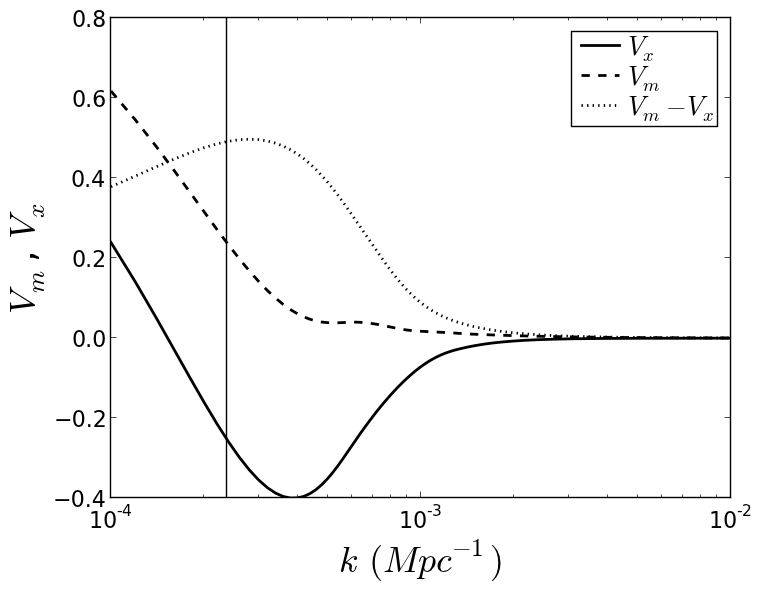}\\[-6mm] \includegraphics[scale=0.4]{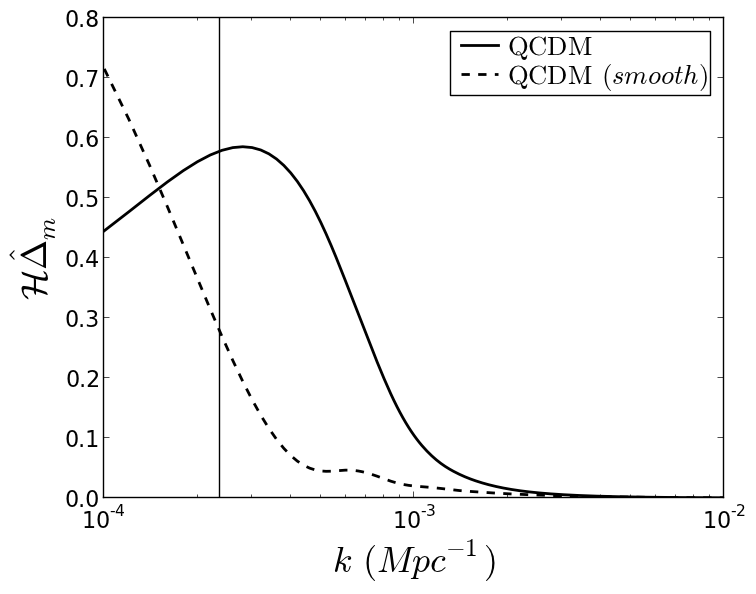} 
\caption{Plots at $z=0$. {\em Top:} The DE and the matter velocity potentials, $V_{x}(k)$ (solid line) and $V_{m}(k)$ (dashed line), respectively; and their difference $V_{m}(k) - V_{x}(k)$ (dotted line). {\em Bottom:} The parameter ${\cal H}\hat{\Delta}_m(k)$, for: clustering QCDM (solid line), and unphysical smooth QCDM (dashed line). The vertical line denotes the (Hubble) horizon.}\label{fig:1}%
\end{figure}%

However (for completeness), if we consider a (generic) clustering DE, \ie\ with $V_x \neq 0$ and $\Delta_x \neq 0 \neq \delta_x$, we get $\Omega_m\hat{\Delta}_m + \Omega_x\hat{\Delta}_x = 0$. It is a straightforward thing to show that, by taking the time derivative of Eq.~\eqref{Poisson} and applying the appropriate equations, we obtain Eq.~\eqref{Phiprime}. Hence, a clustering DE rightly upholds the consistency of the system of equations of general relativity. But as previously mentioned, ${\cal H}\hat{\Delta}_m$ (for example) is physical and will contribute to the matter density perturbation $\Delta_m$ via Eq.~\eqref{DmEvoln}. Generally, for the clustering DE, the growth of ${\cal H}\hat{\Delta}_m$ on super-horizon scales will be restrained by the relative velocity potential, $V_m - V_x$, while for the (unphysical) smooth DE this term grows almost linearly, driven by $V_m$ (see Eq.~\eqref{DmHat}).

We illustrate these behaviours at the present epoch in \fig~\ref{fig:1}, using QCDM. These behaviours explain the matter power spectrum reported in~\cite{Duniya:2013eta}, where it is shown that a smooth QCDM specified by Eq.~\eqref{falseHomDE}, leads to a false (unphysical) amplification of the linear matter power spectrum on super-horizon scales. Note that, although the effect of Eq. \eqref{falseHomDE} may not appear to be significant on sub-horizon scales, it is nevertheless crucial to use the correct and consistent general relativistic equations for (generally) valid analyses.

Henceforth, we reserve the name `smooth' for the unphysical DE, defined by Eq.~\eqref{falseHomDE}.

\subsection{\bf True homogeneous dark energy}\label{subsec:homDE2}  
A physical, consistent homogeneous DE should be one such that it maintains the consistency of the equations of general relativity, by upholding Eq.~\eqref{Consistency}, irrespective of the nature of its equation of state parameter $w_x$.

Given that the entropy (perturbation) of any fluid measures the degree of disorderliness in the fluid, then homogeneity or inhomogeneity of the fluid may suitably be defined with respect to its entropy. Particularly, that the (net) intrinsic entropy perturbation of the fluid vanishes. By the vanishing of the intrinsic entropy perturbation, it implies that the {\em net internal distortion} (\ie~the total change in the inherent distortions -- quantified by the brackets in Eq. \eqref{EntroPert}) of the fluid becomes zero. This way, the fluid may be thought to be constituted by an even distribution of equi-amplitude distortions; hence the fluid is homogeneous (or uniform).

Therefore, here `homogeneity' refers to `uniformity', so that a `true' homogeneous DE is not one entirely devoid of perturbations, but one made up of uniformly distributed (equi-amplitude) perturbations. Thus when the DE intrinsic entropy perturbation vanishes, \ie~$\delta{s}_x = 0$, Eq.~\eqref{EntroPert2} yields
\beq\label{vanishEntrop}
(c^2_{sx} - c^2_{ax})\Delta_x = 0,
\eeq
where either ({\em 1}) $c^2_{ax} = c^2_{sx}$, or ({\em 2}) $\Delta_x = 0$. It is important to note that the definition~\eqref{vanishEntrop} is independent of the choice of the spacetime gauge. It may also be pointed out that, if initially by a priori assumptions $c^2_{ax}=c^2_{sx}$, then automatically $\delta{s}_x=0$; however the converse is not necessarily true: if initially by a priori assumptions $\delta{s}_x=0$ then it may or may not mean that $c^2_{ax}=c^2_{sx}$, since it can also mean that $\Delta_x = 0$ instead.

In what that follows, we investigate the two cases ({\em 1}) and  ({\em 2}), given above.

\subsubsection*{Case~1:~$c^2_{ax} = c^2_{sx}$}\label{subsub:c2ac2s}%
If the adiabatic sound speed is equal to the physical sound speed, \ie~$c^2_{ax} = c^2_{sx}$, then Eq.~\eqref{EoSEvoln} implies that
\beq\label{cs2ca2}
c^2_{ax} \;=\; w_x - \dfrac{w_x'}{3{\cal H}(1+w_x)} \;=\; c^2_{sx}\; \geq\; 0,
\eeq
where this means $3{\cal H} w_x \,\geq\, w_x'/(1+w_x)$, and either: 
\begin{eqnarray*}
\text{(i)} &&w_x > -1\ \text{and}\ w_x' < 0,\ \text{or} \\
\text{(ii)} &&w_x < -1\ \text{and}\ w_x' > 0.
\end{eqnarray*}
Thus, unless $w_x \geq 0$, $w_x$ cannot be an absolute constant. It may only be asymptotic to a fix value, such that $w_x' \neq 0$, otherwise $c_{sx}$ becomes imaginary -- and small perturbations become unstable. Hence, the given homogeneous DE does not admit $w_x={\rm constant} < 0$ (\ie~negative constants). Moreover, conditions (i) and (ii) above, imply that $w_x$ can not oscillate: it may only be either monotonically decreasing ($w_x' < 0$) or monotonically increasing ($w_x' > 0$).  Thus, {\em Case~1}~\eqref{cs2ca2} essentially `fixes' the DE background evolutions.

\begin{figure}\centering
\includegraphics[scale=0.4]{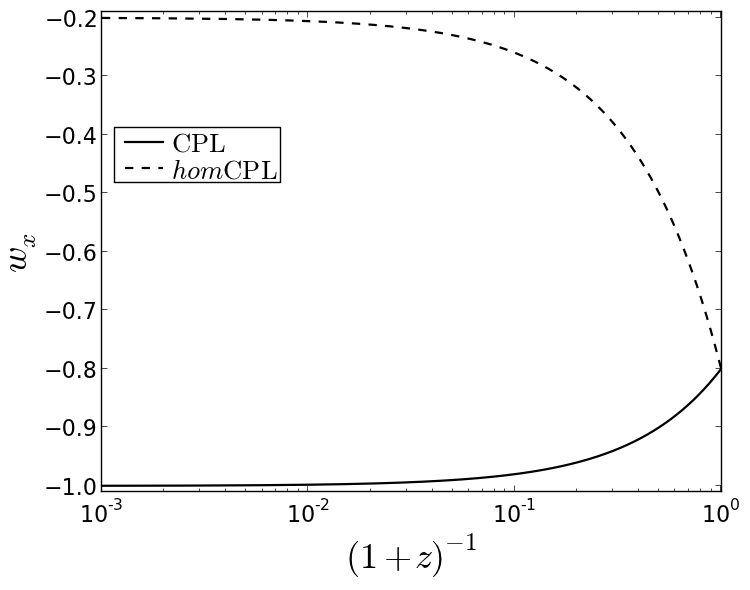} \\ \includegraphics[scale=0.4]{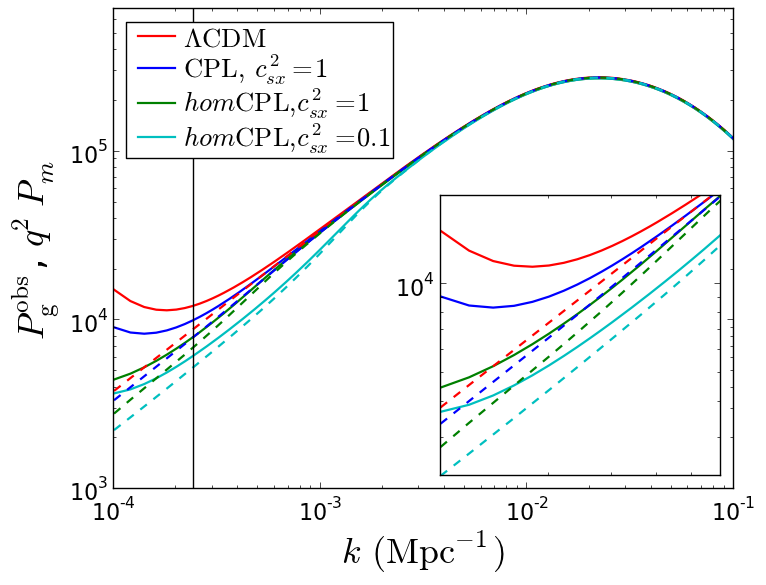}
\caption{{\em Top panel:} The evolution of the equation of state parameter $w_x$, for the CPL~\eqref{CPL:EoS1} (solid black) and the $hom$CPL~\eqref{CPL:EoS2} (dashed black). {\em Bottom panel} (at $z=0$): The radial (\ie~$\mu=1$) galaxy power spectrum $P^{\rm obs}_{\rm g}$ (solid lines) with galaxy bias $b=1$, magnification bias ${\cal Q}=1$; and the matter power spectrum -- given by $q^2P_m$ (dashed lines); $q=\sqrt{2.1}$.}\label{fig:2}
\end{figure}

To illustrate {\em Case~1}~\eqref{cs2ca2}, we consider the well known Chevallier-Polarski-Linder (CPL) parametrization \cite{Chevallier:2000qy,Linder:2002et}:
\beq\label{CPL:EoS}
w_x(a) \;=\; w_0 +w_a(1-a),
\eeq
where the scale factor $a = (1+z)^{-1}$, with $z$ being the redshift; $w_0$ and $w_a$ are (free) constants. We consider two scenarios of $w_x$: a generic clustering DE (CPL) and a homogeneous DE ($hom$CPL), given by
\bea\label{CPL:EoS1}
{\rm CPL:} && w_x \geq -1,~w_x'\geq 0,~\Delta_x \neq 0, \\ \label{CPL:EoS2}
hom{\rm CPL:} && w_x > -1,~w_x'<0,~\Delta_x \neq 0,
\eea
where we choose $w_0=-0.8$ and $w_a=-0.2$ for CPL; $w_0=-0.8$ and $w_a=0.6$ for $hom$CPL, \ie~here we consider only the scenario (i) of {\em Case~1}~\eqref{cs2ca2}. We show the behaviour of $w_x$ in \fig~\ref{fig:2} (top panel), for both the CPL \eqref{CPL:EoS1} and the $hom$CPL \eqref{CPL:EoS2}, respectively.

Throughout this work, we initialize evolutions at the decoupling epoch, given by $1+z_d = 10^3$. We normalize all the power spectra on small scales, at $z=0$: \ie~by choosing the same matter density parameter $\Omega_{m0}=0.24$ and Hubble constant $H_0=73~\kms\mpc$ for all cases. Thus all of the power spectra match each other at $z=0$, on small scales. (The advantage of this is that any clustering or GR effects become isolated on large scales.) We used adiabatic initial conditions (see Appendix~\ref{sec:AICs}) for the perturbations.

We show in \fig~\ref{fig:2} (bottom panel), the radial galaxy power spectrum $P^{\rm obs}_{\rm g}$ (see Appendix~\ref{sec:PowerSpec}) with galaxy bias~\cite{Challinor:2011bk}--\cite{Baldauf:2011bh} $b=1$ and magnification bias~\cite{Jeong:2011as} ${\cal Q}=1$; and the matter power spectrum $P_m$: at the present epoch, \ie~$z=0$. By our normalization, we see that $P^{\rm obs}_{\rm g}$ can be approximated on sub-Horizon scales by $q^2P_m$, with $q=\sqrt{2.1}$. Moreover, although DE clusters in the CPL, we see that it rather leads to higher power on horizon scales in both $P^{\rm obs}_{\rm g}$ and $P_m$, \ie~relative to the $hom$CPL for equal values of $c^2_{sx}$ (see~\cite{Creminelli:2008wc,Creminelli:2009mu,Takada:2006xs,Bean:2003fb,dePutter:2010vy}--\cite{Mehrabi:2015hva}, for the effects of $c^2_{sx}$). This may be owing to the behaviour of $w_x$ for $hom$CPL, which suggests that DE sets in relatively earlier for the $hom$CPL -- hence causing the matter perturbations to have less time to cluster, thereby resulting in relatively lower power spectra. Obviously, {\em Case~1}~\eqref{cs2ca2} implies that for equal values of $c^2_{sx}$, the difference between a homogeneous DE and a clustering DE is mainly governed by the background, with little to do with the perturbations. However, one may expect that this difference strongly pertains perturbations, and that a homogeneous DE results in higher power spectra on large scales, relative to a clustering DE -- given that the perturbative effect of the homogeneous DE should be negligible (or even absent).

We also observe the dependence of $hom$CPL on $c^2_{sx}$, \ie~in the power spectra, with smaller values of $c^2_{sx}$ resulting in more power suppression -- since in which case the DE perturbations are able to cluster earlier and on smaller scales; thus suppressing most of the matter growth. However, we see the effect of the general relativistic (GR) corrections~\cite{Duniya:2013eta,Challinor:2011bk}--\cite{Baldauf:2011bh}, \cite{Bonvin:2011bg}--\cite{Bonvin:2014owa} in $P^{\rm obs}_{\rm g}$: they lead to a sizeable power boost (relative to $P_m$) on horizon scales. Moreover, we observe that the GR corrections also result in significant differentiation of the given DE scenarios. 

Nevertheless, for self-consistent models, \eg~the QCDM (specified by a scalar field $\varphi$) which evolve along a potential given (generically) by $U(\varphi) \neq {\rm constant}$, we have
\beq\label{cs2ca2Quint}
c^2_{ax} = 1 +\dfrac{2a^2 U_{|\varphi}}{3{\cal H}\bar{\varphi}'} \;\neq\; c^2_{sx} = 1,
\eeq
where $U_{|\varphi} \equiv \partial U(\varphi) / \partial\bar{\varphi}$, with $c^2_{ax}$ and $c^2_{sx}$ being as defined in sections~\ref{sec:bkgnd} and~\ref{sec:Entropy}, respectively. Thus by Eq.~\eqref{cs2ca2Quint}, $c^2_{ax} \neq c^2_{sx}$, which then disallows {\em Case~1}~\eqref{cs2ca2}. Therefore, $w_x$ may oscillate (see, \eg~\cite{Bassett:2007aw,Okouma:2012dy,Shafieloo:2014ypa}) or take any behaviour. However, one may choose to fix $U(\varphi) = {\rm constant}$, thereby satisfying {\em Case~1}~\eqref{cs2ca2} (in principle). Practically though, this choice leads to $w_{x}$ violating {\em Case~1}~\eqref{cs2ca2}, \ie~by becoming $w_{x}=-1$ for $0 \leq z \lesssim 100$, in which case the perturbations equations become unsolvable numerically. Thus {\em Case~1}~\eqref{cs2ca2} is `impractical' for the QCDM. 

In general, {\em Case~1}~\eqref{cs2ca2} is unsatisfactory, given that it still depends on the behaviour  or choice of $c^2_{sx} \geq 0$.

\subsubsection*{Case~2:~$\Delta_x = 0$}
On the other hand, if the DE comoving density perturbation vanishes, \ie~$\Delta_x = \delta_x - 3{\cal H}(1+w_x)V_x = 0$, then it implies that the correction to Eq.~\eqref{falseHomDE} is suitably given by
\beq\label{vanishDelta}%
\delta_x = 3{\cal H}(1+w_x)V_x,\quad V_x\neq 0,
\eeq
where consequently, $\Delta'_x=0$. Thus Eq. \eqref{vanishDelta} implies that the homogeneous DE should not have any density perturbations $\Delta_x$ in comoving gauge, but may posses the fractional density fluctuations $\delta_x$ which generate peculiar velocities with potentials $V_x$. 

Note that given Eq. \eqref{comvD}, {\em Case~2} \eqref{vanishDelta} implies that a homogeneous DE has zero density perturbation in its rest frame, \ie~$\delta\rho_x|_{\rm rf} = \bar{\rho}_x\Delta_x =0$. This is the physical statement of {\em Case~2} \eqref{vanishDelta}, which is a crucial statement -- as it suggests that DE homogeneity or inhomogeneity should be defined relative to the DE rest frame, \ie~by whether or not its density perturbations vanish in its rest frame.

In fact, {\em Case~2}~\eqref{vanishDelta} readily holds for the QCDM when $\delta{s}_x=0$ (\ie~from Eq.~\eqref{EntroPert}); in which case we have
\beq\label{homQuint}
c^2_{ax}\delta{\rho}_{\varphi} \;=\; \delta{p}_{\varphi} \,=\, \delta{\rho}_{\varphi} + 3{\cal H}(1+w_x)(c^2_{ax} - 1)\bar{\rho}_{\varphi}V_x,
\eeq
where by collecting terms with $\delta{\rho}_{\varphi}$ to one side and dividing through by $(c^2_{ax} - 1)\bar{\rho}_{\varphi}$, we get $\delta_x = 3{\cal H}(1+w_x)V_x$. We have used Eq.~\eqref{cs2ca2Quint}; $\bar{\varphi}'^2 = a^2(1+w_x)\bar{\rho}_{\varphi}$, $w_x=\bar{p}_{\varphi}/\bar{\rho}_{\varphi}$ with the perturbations $V_x = -\delta{\varphi}/\bar{\varphi}'$, $\delta_x = \delta\rho_{\varphi}/\bar{\rho}_{\varphi}$ and
\bea\label{pQuintPert}
\delta{p}_{\varphi} &=& \delta{\rho}_{\varphi} - 2U_{|\varphi}\delta{\varphi},\\ \label{rhoQuintPert}
\delta{\rho}_{\varphi} &=& a^{-2}\bar{\varphi}'\left(\delta{\varphi}' - \bar{\varphi}'\Phi'\right) + U_{|\varphi}\delta{\varphi}.
\eea 
Besides, by considering the definition of the intrinsic entropy perturbation specifically for QCDM given by $\Gamma$ \cite{Bartolo:2003ad}, which relates to the entropy perturbation of section~\ref{sec:Entropy} by $\Gamma = w_x\, \delta{s}_x / (c^2_{sx} -c^2_{ax}) = \Delta_x$ -- where the second equality follows from Eq. \eqref{EntroPert2}, then it automatically follows that $\Delta_x=0$ when we set $\Gamma=0$ (which is given therein as the adiabaticity condition for quintessence). Thus a clustering quintessence will have $\Delta_x \neq 0$ and a homogeneous quintessence will have $\Delta_x = 0$: both scenarios having the same (background) equation of state parameter $w_x = (\bar{\varphi}'^2 - 2a^2U) / (\bar{\varphi}'^2 + 2a^2U)$, which varies (in general) by $-1 \leq w_x \leq 1$. 

It should be pointed out that the discussion on Eqs. \eqref{homQuint}--\eqref{rhoQuintPert} is not implying that QCDM models are {\it generically} in the form of {\em Case~2} \eqref{vanishDelta}, nor are we claiming that there is any other particular DE model (or class of models) that {\it naturally} exists in such form. Instead, {\em Case~2} \eqref{vanishDelta} is rather a proposition for a `general' homogeneity condition which may be applied to any (dynamical) DE model. (Notice that {\em Case~2} \eqref{vanishDelta} was arrived at -- mainly via Eq. \eqref{vanishEntrop} -- without assuming any DE models). Hence the QCDM models are only used as a reference or an example of a particular, well known DE model that {\em Case~2} \eqref{vanishDelta} may easily be applied to. Moreover, {\em Case~2} \eqref{vanishDelta} applies to the CPL parametrization -- which approximates (canonical) scalar-field DE models.

In general, {\em Case~2}~\eqref{vanishDelta} suggests that DE may be homogeneous only when the observer is comoving with the source, irrespective of the DE background specifications. Thus a homogeneous DE may have perturbations, but only such that these perturbations combine to cancel out in comoving gauge (and hence, in the DE rest frame). Moreover, the matter power spectrum physically makes sense only when computed in comoving gauge, since otherwise, it becomes gauge-dependent and varies with the observers on large scales (see \eg~\cite{Bonvin:2011bg,Flender:2012nq}). Thus {\em Case~2}~\eqref{vanishDelta}, being defined in comoving gauge, can lead to (physical) observable implications in the power spectrum. Moreover, the effect of {\em Case~2} \eqref{vanishDelta} on the matter perturbations will be imposed directly, rather than indirectly via the background evolutions -- as in {\em Case~1} \eqref{cs2ca2}. This way, the imprint of the given homogeneous DE will bear directly on the growth of structure. 

An important advantage of {\em Case~2}~\eqref{vanishDelta} over {\em Case~1} \eqref{cs2ca2} is that, unlike {\em Case~1}~\eqref{cs2ca2}, {\em Case~2}~\eqref{vanishDelta} permits an arbitrary background behaviour for the given DE: $w_x$ may take any nature (constant or otherwise). This is important for models with either oscillatory or constant $w_x$. A further advantage of {\em Case~2}~\eqref{vanishDelta} is that it eliminates the dependence of the perturbations on $c^2_{sx}$ (via Eq.~\eqref{VxEvoln}), given that $\Delta_x=0$ or $\delta_x=3{\cal H}(1+w_x)V_x$. Hence the given homogeneous DE is completely independent of the choice or nature of $c^2_{sx}$ (just like in \LCDM). This further removes the risk that accompanies a bad choice or wrong modelling of $c^2_{sx}$. Moreover, it also reduces the parameter space that needs to be constrained. 

\begin{figure}\centering
\includegraphics[scale=0.4]{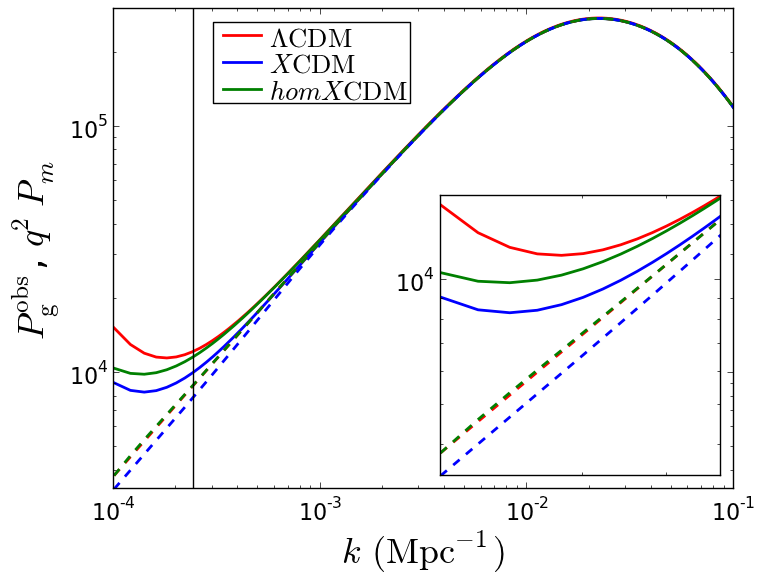}
\caption{Plots at $z=0$: The radial (\ie~$\mu=1$) galaxy power spectrum $P^{\rm obs}_{\rm g}$ (solid lines) with galaxy bias $b=1$ and magnification bias ${\cal Q}=1$; and the matter power spectrum -- given by $q^2P_m$ (dashed lines) with $q=\sqrt{2.1}$. The vertical line denotes the horizon at the given redshift.}\label{fig:3}
\end{figure}

To illustrate {\em Case~2}~\eqref{vanishDelta}, we use only the generalized phenomenological model \eqref{CPL:EoS} (ignoring the particular scenario of the QCDM). Hereafter we denote clustering DE by $X$CDM -- given by Eq. \eqref{CPL:EoS1}, and denote the associated homogeneous DE by $homX$CDM -- which has the same background parameters as $X$CDM, but with its perturbations being specified by {\em Case~2} \eqref{vanishDelta}. We use $c_{sx}=1$ for all numerical computations (\ie~for $X$CDM).

We show in \fig~\ref{fig:3} the galaxy power spectrum $P^{\rm obs}_{\rm g}$ with $b=1$ and ${\cal Q}=1$, and the associated matter power spectrum $P_m$: for $X$CDM and $homX$CDM, at $z=0$. We see that on sub-horizon scales, $P^{\rm obs}_{\rm g}$ can be approximated by $q^2P_m$, where $q=\sqrt{2.1}$. Moreover, unlike the results by {\em Case~1}~\eqref{cs2ca2} (see \fig~\ref{fig:2}), where the clustering DE results in large-scale boost in the power spectra relative to the homogeneous DE, here we see that although $c^2_{sx}=1$ for the $X$CDM, we get large-scale power suppression in both $P^{\rm obs}_{\rm g}$ and $P_m$ relative to those for $homX$CDM (and \LCDM) -- \ie~the power spectra for $X$CDM are lower than those for $homX$CDM, on horizon scales. This implies that on horizon scales, the effect of $c^2_{sx}$ in $X$CDM is less significant and hence the DE perturbations are able to cluster enough to suppress the growth of the matter perturbations. On the other hand, the DE density perturbations vanish on all scales for $homX$CDM (in comoving gauge); thus the matter perturbations are able to grow more. Consequently, we get the relative boost in the power spectra in $homX$CDM. 

Moreover, we see that the \LCDM~gives a sizeable deviation in $P^{\rm obs}_{\rm g}$ relative to $homX$CDM, on super-horizon scales. (Note however that in reality, this statement depends on: ($1$) the cosmic variance on these scales, and ($2$) the error bars achievable by a given survey experiment. But for the purpose of this work, we leave out (throughout this work) any exact experimental aspects.) This deviation illuminates the sensitivity of GR corrections to changes in $w_x$; this sensitivity will be crucial in discriminating the \LCDM\ from a dynamical homogeneous DE model -- with the future large scale surveys. We also observe that, $P_m$ of $homX$CDM is identical to that of \LCDM~on all scales. This reveals that, the linear matter power spectrum is incapable of distinguishing a dynamical homogeneous DE (given by {\em Case~2}~\eqref{vanishDelta}) from \LCDM, on large scales -- when their power spectra are normalized on small scales (at the given epoch).

In the intensity mapping of neutral hydrogen (HI), usually the individual radio sources are not counted: only the diffuse 21 cm line emission of a number of sources is detected \cite{Hall:2012wd}. Then given that the volume distortion in the observed density perturbation mainly leads to the `amplification' of the number of sources, it is therefore taken that the volume distortion does not contribute in HI intensity mapping: hence, the magnification bias is often set to ${\cal Q}=1$, which results in the volume distortion being cancelled out in the observed density perturbation. (Pure galaxy number count surveys -- in which mainly individual galaxies are counted -- correspond to ${\cal Q}=0$: thus eliminating the cosmic magnification of the galaxies.)

\begin{figure}\centering
\includegraphics[scale=0.4]{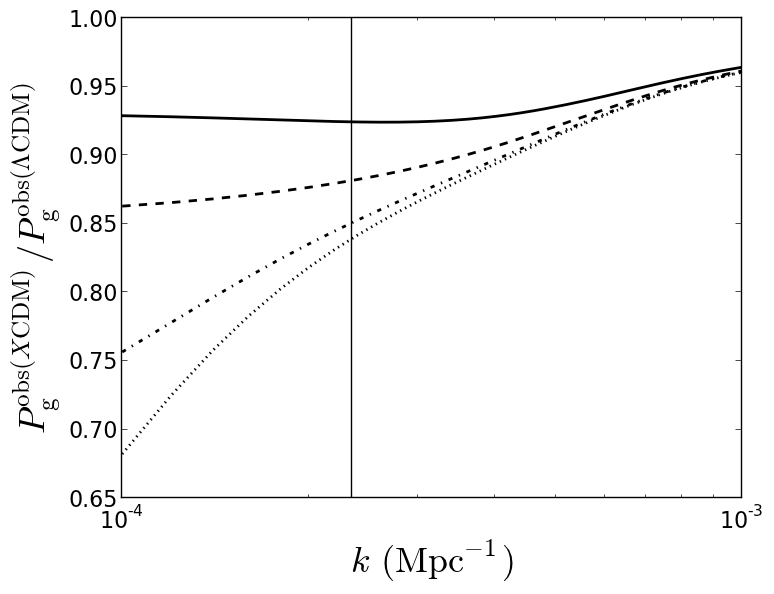} \\[-6mm]
\includegraphics[scale=0.4]{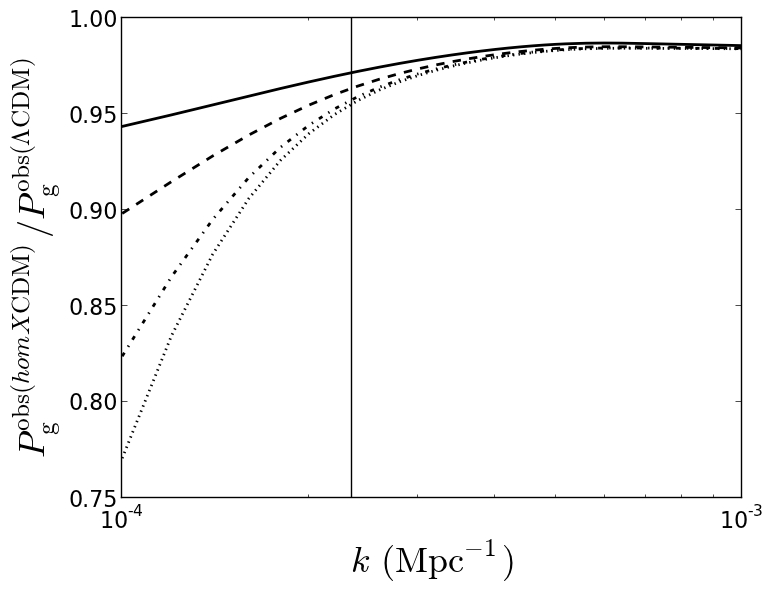} \\[-6mm]
\includegraphics[scale=0.4]{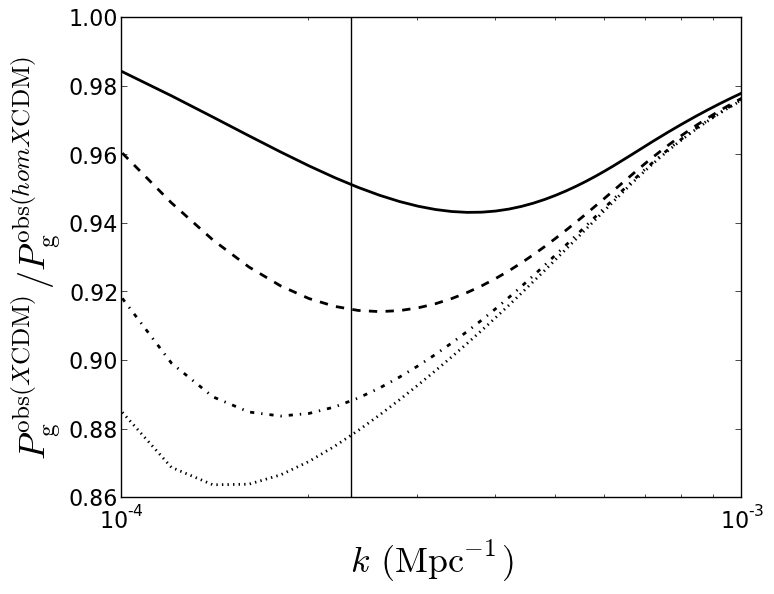} 
\caption{The plots of the ratios of $P^{\rm obs}_{\rm g}$ (with $\mu=1$) at $z=0.1$, with galaxy bias $b=1$. The panels show the ratios for: $X$CDM relative to \LCDM\ (top), $homX$CDM relative to \LCDM\ (middle) and $X$CDM relative to $homX$CDM (bottom).  The lines denote: ${\cal Q}=-1$ (solid), ${\cal Q}=0.5$ (dashed), ${\cal Q}=0.9$ (dot-dashed) and ${\cal Q}=1$ (dotted). The vertical line denotes the horizon at the given redshift.}\label{fig:4}
\end{figure}

In \fig~\ref{fig:4} we illustrate the effect of the magnification bias ${\cal Q}$. We show the ratios of the radial (\ie~with $\mu=1$) galaxy power spectrum $P^{\rm obs}_{\rm g}$ at the epoch $z=0.1$, with galaxy bias $b=1$: for ${\cal Q}=-1,~0.5,~0.9,~1$. In the top panel, we give the ratios for $X$CDM relative to \LCDM. Obviously, we see that the ratio of the two models varies with different values of the magnification bias -- which is not surprising since the relation between the magnification bias and the power spectrum is non-trivial (see Eqs.\eqref{Pk:Obs}--\eqref{calB}). If, for example, ${\cal Q}$ had a simple scaling relation to $P^{\rm obs}_{\rm g}$, \ie~if ${\cal Q}$ appeared in $P^{\rm obs}_{\rm g}$ merely as a multiplicative factor, then the ratio between the given models will remain the same for all values of ${\cal Q}$. Moreover, the matter density perturbation and velocity potential, and the gravitational potential (hence, the GR corrections) in $X$CDM will inherently differ from those in \LCDM: owing to the difference in the DE perturbations, and the DE equation of state parameter. At low $z$, the matter overdensity will be larger in \LCDM\ than in $X$CDM on horizon scales. Consequently, for a given ${\cal Q}$, the amplitude of $P^{\rm obs}_{\rm g}$ is larger in \LCDM\ than in $X$CDM -- thus the ratio of the two models is less than unity (on horizon scales), as seen in the plots.

We see that the given models start to differentiate from each other on horizon scales, whereas on scales well within the horizon they match each other -- which is mainly owing to our normalization (at $z=0$). Furthermore, we observe that $P^{\rm obs}_{\rm g}$ in $X$CDM is consistently suppressed relative to that of \LCDM~as the magnification bias increases: from ${\cal Q}=-1$ to ${\cal Q}=1$. This implies that the value of the magnification bias can be crucial in differentiating a clustering DE from the cosmological constant (at the given $z$), given that the ratio between the models is sensitive to the values of ${\cal Q}$. Moreover, note that for a given value of ${\cal Q}$, the large-scale amplitude of the power spectrum will be higher in \LCDM\ than in $X$CDM (as previously explained): $\Lambda$ has no perturbations to suppress the matter perturbations; moreover, the equation of state parameter of $\Lambda$ is weaker than $w_{x}$ (for dynamical DE) -- noting that the bigger the equation of state parameter, the stronger the DE.

\fig~\ref{fig:4} (middle panel) also shows the ratio of $P^{\rm obs}_{\rm g}$, for $homX$CDM relative to that of \LCDM: for the given values of ${\cal Q}$. The results are similar to those between $X$CDM and \LCDM\ (in the top panel), except that the amplitude of the ratios in the middle panel are relatively higher -- for a given value of ${\cal Q}$. This is mainly a clustering effect of DE: unlike in $homX$CDM, the large-scale clustering of DE in $X$CDM leads to the suppression of the matter perturbations (and thus, GR corrections) -- for the same $w_x$; hence resulting in a relatively lower galaxy power prediction in the $X$CDM, on the given scales. Moreover, just as in the top panel, the ratios of $P^{\rm obs}_{\rm g}$ for $homX$CDM relative to \LCDM\ also suggest that the value of ${\cal Q}$ can be crucial in differentiating a homogeneous DE from the cosmological constant.

The bottom panel of \fig~\ref{fig:4} shows the ratios of $P^{\rm obs}_{\rm g}$: for $X$CDM relative to $homX$CDM. Similarly, we see that as ${\cal Q}$ increases, $X$CDM becomes consistently suppressed relative to $homX$CDM, \ie~the bigger the value of the magnification bias, the wider the difference between the clustering DE and the homogeneous DE (and vice versa). This suggests that future surveys that depend on cosmic magnification -- \eg~the HI intensity mapping (see \eg~\cite{Duniya:2013eta,Hall:2012wd,Maartens:2012rh}, which correspond to ${\cal Q}=1$) -- will be useful in distinguishing or identifying a homogeneous DE from a clustering DE in the large scale analysis, particularly on horizon scales. Moreover, we observe that the ratios (in the three panels) grow with decreasing ${\cal Q}$, suggesting that at a particular $z$, GR effects become enhanced as cosmic magnification bias decreases. 

Note that, as previously mentioned, the various ratios in the top and the middle panels (\fig~\ref{fig:4}), respectively, inherently contain the effect of the respective equation of state parameters of the given models -- \ie~for a value of ${\cal Q}$, the difference between successive ratios is not only owing to the effect of the perturbations, but also owing to the background difference of the models (via $w_x$). However, in the bottom panel, the ratios are resulting mainly from the difference in the perturbations of the models -- in the comoving gauge. The imprint of $w_x$ is effectively factored out, given that we use the same $w_x$ for both $X$CDM and $homX$CDM; thus revealing mainly the consequence of the {\em Case~2} \eqref{vanishDelta}.

\begin{figure}\centering
\includegraphics[scale=0.38]{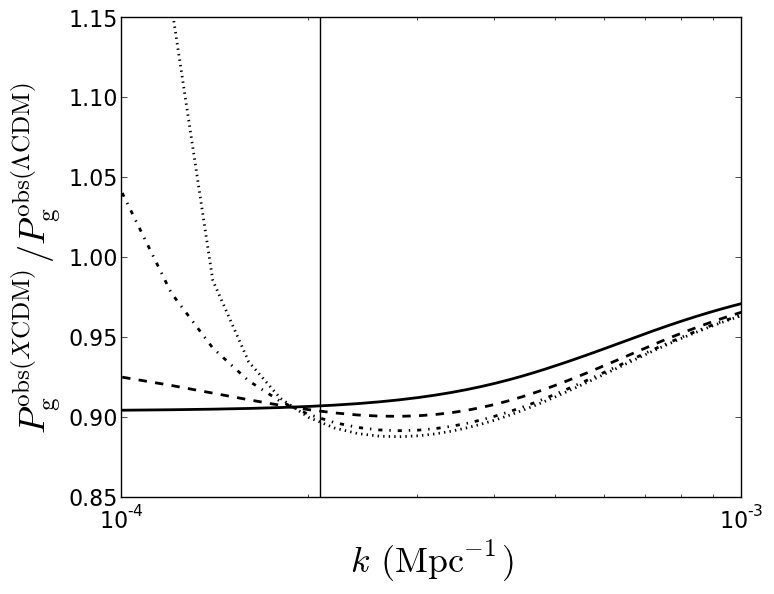} \\[-6mm]
\includegraphics[scale=0.38]{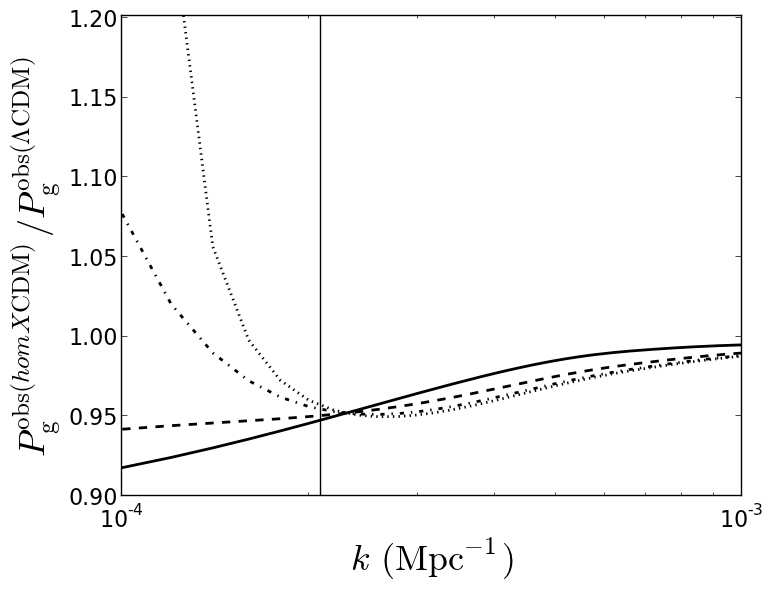} \\[-6mm]
\includegraphics[scale=0.38]{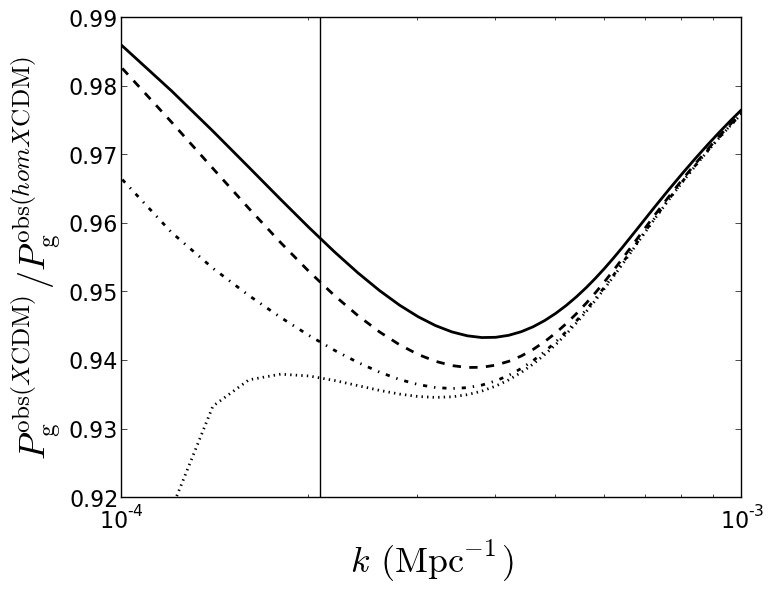} 
\caption{The plots of the ratios of $P^{\rm obs}_{\rm g}$ (with $\mu=1$) at $z=1$, with galaxy bias $b=1$. Line notations are as in \fig~\ref{fig:4}.}\label{fig:5}
\end{figure}

In \fig~\ref{fig:5}, we repeat the plots of \fig~\ref{fig:4}, but here at $z=1$. We observe that, generally (at the given $z$), on scales near the horizon the ratios maintain a consistent decrease with increasing ${\cal Q}$. However, in the top and the middle panels, respectively, the ratios grow on super-horizon scales: with the ratio for ${\cal Q}=-1$ being the lowest, and that for ${\cal Q}=1$ being the highest. This may be attributed to the magnification effect (\ie~terms proportional to ${\cal Q}$, in $P^{\rm obs}_{\rm g}$): which increases on large scales at $z \gtrsim 1$, being boosted by stronger $w_x > -1$. (The clustering effect of DE generally diminishes with increasing $z$ -- hence at high $z$ the DE effect is mainly governed by the (background) equation of state parameter.) Thus $P^{\rm obs}_{\rm g}$ becomes enhanced for increasing ${\cal Q}$. However, in the bottom panel the effect of $w_x$ is factored out, and only the effect of ${\cal Q}$ is seen. Moreover, Figs.~\ref{fig:4} and \ref{fig:5} generally show that for a given magnification bias, the ratios are higher on super-horizon scales at $z=1$ than at $z=0.1$; hence implying higher GR effects at $z \gtrsim 1$.

In general, unlike {\em Case~1}~\eqref{cs2ca2} which further depends on the perturbation modes being well within the horizon and $c^2_{sx} \simeq 1$, {\em Case~2}~\eqref{vanishDelta} is a definitive condition for DE homogeneity: once it is chosen, no other requirements are needed. Moreover, apart from solving the problem posed by Eq.~\eqref{falseHomDE} and the elimination of $c^2_{sx}$ from the equations, {\em Case~2} \eqref{vanishDelta} can also conveniently admit particularly $w_x=-1$ while still allowing the DE perturbations. Thus {\em Case~2}~\eqref{vanishDelta} is more robust, and is hereby considered as the right DE homogeneity condition.

\section{Conclusion}\label{sec:Concln}
We have shown analytically that the DE homogeneity assumption $\delta_x=0=V_x$ (with the evolution equations being discarded) violates the self-consistency of the equations of general relativity. We showed in Newtonian gauge that, unless the equation of state parameter of the given homogeneous DE is strictly $w_x = -1$, this assumption introduces a contradiction in the equations of general relativity. In essence, provided $w_x \neq -1$, DE must have perturbations.

We have proposed a correct homogeneity condition for DE, which is valid irrespective of the nature of the DE equation of state parameter or spacetime gauge -- by supposing that the DE intrinsic entropy perturbation vanishes: which leads to the vanishing of the DE overdensity in comoving gauge (and hence, in the DE rest frame). Thus we correct the wrong homogeneity assumption ($\delta_x=0=V_x$) by the following: $\delta_x=3{\cal H}(1+w_x)V_x$, with $V_x \neq 0$. A homogeneous DE hence, is given not as one devoid of perturbations, but rather as one with vanishing density perturbations in comoving gauge -- \ie~one with zero density perturbations in its rest frame. Such kind of DE is not impractical.

Using a phenomenological DE model, we investigated the consequence of our approach in the observed galaxy power spectrum. By normalizing the models at the present epoch on small scales, we found that: a clustering DE, a homogeneous (dynamical) DE and the cosmological constant, are each distinguishable from the others in the observed galaxy power spectrum, on horizon scales -- suitably for high cosmic magnification bias. 

Moreover, the results show that for a given magnification bias, GR effects in the galaxy power spectrum become enhanced with increasing redshift.

\[ \] {\bf Acknowledgements:} Thanks to Roy Maartens, Bruce Bassett, Eric Linder, David Polarski and Kazuya Koyama, for useful comments. This work was supported by the South African Square Kilometre Array Project, the South African National Research Foundation and by a Royal Society (UK)/ National Research Foundation (SA) exchange grant.

\appendix %

\section{Adiabatic initial conditions}\label{sec:AICs}%
All evolutions in this work are initialized at the photon decoupling epoch, $z=z_d$. We use adiabatic initial conditions, which follow from the vanishing of the relative entropy perturbation~\cite{Duniya:2013eta,Bartolo:2003ad,Kodama:1985bj}--\cite{Malik:2004tf}, given at $z_d$ by
\beq\label{adic}
\delta_x = \left(1 +w_x\right)\delta_m.
\eeq
By using that at $z_d$, we have
\beq
V_x = V_m, 
\eeq
then we obtain, \ie~given \eqref{adic}, that
\beq
\Delta_x = (1 +w_x)\Delta_m.
\eeq
These equations together with the Einstein de Sitter initial condition $\Phi'(z_d)=0$,  lead to the initial perturbations
\bea\label{Delta_mi}
\Delta_m(k) &=& \dfrac{-2k^2}{3\left(1 + \Omega_x w_x\right){\cal H}^2}  \Phi_d(k), \quad\\ \label{Delta_xi}
V_m(k) &=& \dfrac{-2}{3\left(1 + \Omega_x w_x\right){\cal H}} \Phi_d(k),
\eea\\
where we take $\Phi_d=\Phi(z_d)$ as given by~\cite{Duniya:2013eta}.

\section{The galaxy power spectrum}\label{sec:PowerSpec}
In order to adequately account for the correct galaxy distribution on large scales, we use the observed galaxy density perturbation \cite{Duniya:2013eta,Challinor:2011bk,Jeong:2011as,Yoo:2010ni,Bonvin:2011bg,Yoo:2014kpa} to compute that galaxy power spectrum $P^{\rm obs}_{\rm g}$, which is approximated in the flat-sky limit (in Fourier space) by~\cite{Duniya:2015nva,Jeong:2011as}
\bea\label{Pk:Obs}
{P^{\rm obs}_{\rm g} \over P_m} = \left(b + f\mu^2\right)^2 + 2\left(b + f\mu^2\right) {{\cal A} \over x^2} + {{\cal A}^2 \over x^4} +\mu^2\dfrac{{\cal B}^2}{x^2}, 
\eea
where $P_m$ is the matter power spectrum \cite{Duniya:2013eta,Duniya:2015nva}; $\mu$ is the cosine of the angle between the line-of-sight and the wavevector ${\bf k}$, with $k =|{\bf k}|$ being the wavenumber; $x\equiv k/{\cal H}$ is a dimensionless parameter, and 
\bea \label{calA}
{\cal A} &= & x^2 \left[4{\cal Q} - b_e - 1 + \dfrac{{\cal H}'}{{\cal H}^2}  + 2\frac{\left(1-{\cal Q}\right)}{r{\cal H}} + \dfrac{\Phi'}{{\cal H}\Phi}\right] \dfrac{\Phi}{\Delta_m} \nn
&& +\; \left(3-b_e\right) f,\\ \label{calB} 
{\cal B} &= & -\left[ b_e -2{\cal Q}-\frac{\mathcal{H}'}{\mathcal{H}^2}-2\frac{\left(1-{\cal Q}\right)}{r \mathcal{H}} \right] f, \;\;\;
\eea
where $f \equiv k^2V_m / ({\cal H} \Delta_m)$ (see also~\cite{Duniya:2015nva}). Note that in \LCDM, this $f$ reduces to the standard linear growth rate of matter energy density perturbation. We have neglected all integral terms, and assumed constant comoving galaxy number density -- thus the galaxy evolution bias $b_e = 0$; $r$ is the comoving distance at the observed $z$.

%

\end{document}